\documentclass[journal,a4paper,fleqn]{IEEEtran}

\ifCLASSINFOpdf
\else
\fi

\usepackage[square,numbers]{natbib}

\usepackage{algorithm}

\usepackage[noend]{algpseudocode}

\usepackage{array}
\usepackage{amsthm}
\usepackage{amsmath}
\usepackage{hyperref}
\usepackage{graphicx}
\usepackage{tabularx}
\usepackage{multirow}
\usepackage{tabulary,booktabs}
\usepackage[table]{xcolor}
\usepackage{float,subfigure}

\algnewcommand{\Initialize}[1]{%
  \State \textbf{Initialize:}
  \Statex \hspace*{\algorithmicindent}\parbox[t]{.8\linewidth}{\raggedright #1}
}
\algnewcommand\algorithmicswitch{\textbf{switch}}
\algnewcommand\algorithmiccase{\textbf{case}}
\algnewcommand\algorithmicassert{\texttt{assert}}
\algnewcommand\Assert[1]{\State \algorithmicassert(#1)}%

\algdef{SE}[SWITCH]{Switch}{EndSwitch}[1]{\algorithmicswitch\ #1\ \algorithmicdo}{\algorithmicend\ \algorithmicswitch}%
\algdef{SE}[CASE]{Case}{EndCase}[1]{\algorithmiccase\ #1}{\algorithmicend\ \algorithmiccase}%
\algtext*{EndSwitch}%
\algtext*{EndCase}%

\begin{document}

\title{New intelligent defense systems to reduce the risks of Selfish Mining and Double-Spending attacks using Learning Automata}

\author{Seyed~Ardalan~Ghoreishi,
        and~Mohammad~Reza~Meybodi}

\markboth{Smart Defenses against DS and SM attacks}%
{Shell \MakeLowercase{\textit{Ghoreishi and Meybodi.}}: Bare Demo of IEEEtran.cls for IEEE Journals}

\maketitle

\begin{abstract}
In this paper, we address the critical challenges of double-spending and selfish mining attacks in blockchain-based digital currencies. Double-spending is a problem where the same tender is spent multiple times during a digital currency transaction, while selfish mining is an intentional alteration of a blockchain to increase rewards to one miner or a group of miners. We introduce a new attack that combines both these attacks and propose a machine learning-based solution to mitigate the risks associated with them. Specifically, we use the learning automaton, a powerful online learning method, to develop two models, namely the SDTLA and WVBM, which can effectively defend against selfish mining attacks. Our experimental results show that the SDTLA method increases the profitability threshold of selfish mining up to 47$\%$, while the WVBM method performs even better and is very close to the ideal situation where each miner's revenue is proportional to their shared hash processing power. Additionally, we demonstrate that both methods can effectively reduce the risks of double-spending by tuning the $Z$ Parameter. Our findings highlight the potential of SDTLA and WVBM as promising solutions for enhancing the security and efficiency of blockchain networks.
\end{abstract}


\IEEEpeerreviewmaketitle

\section{Introduction}
Blockchain can be considered a distributed ledger where all approved transactions are stored in its blocks \cite{R1}. This chain is constantly growing with the addition of new blocks. For example, a famous blockchain named Bitcoin relies on an incentive mechanism called mining. Every node that participates in mining progress is called a miner. Miners are trying to produce blocks and broadcast them. When a miner mines a block, they will receive rewards. Miners often pool their resources (computing and processing power) to form a mining pool. By doing this, they can mine more blocks and thus share the mining reward. When more than one block extends the previous block, the main chain is determined by a fork-resolving policy. In this case, the miner selects the longest chain, or in the case of multiple chains with the same length, the chain that receives the next block will be considered the main chain. This forked situation is called a block race, and an equal-length block race is called a tie. As long as more than half of the mining power follows the protocol, the probability that a miner will gain the next block reward is equal to the miner’s computing power \cite{R2}. These rules provide the opportunity to launch a selfish mining attack, as offered by Eyal et al. \cite{R3}. Selfish mining refers to the efforts of a destructive miner to increase their share of the mining reward. The attacker hides the mined blocks for a period of time and then releases several blocks immediately, causing the other miners to lose their blocks \cite{R3}.

Double spending is one of the most common attacks that take place by abusing the transaction confirmation mechanism. All transactions on the platform of a blockchain must be approved by other users to be recognized as valid and approved transactions, and of course, this process takes some time; attackers can use this time to take advantage and trick the system into using the same coins in other transactions as well. In general, the chances of double-spending attacks succeeding are low. However, if the attacker continues to try these attacks alternately with the $\alpha$ computing power, he will eventually succeed \cite{R4}. Instead of focusing on the likelihood of success, we should focus on the cost of these attacks. Any failed attempt to double-spending attack would cost the attacker the equivalent of losing the reward he would receive if he honestly released his blocks instead of hiding them. This is where selfish mining can help the double-spending attacker. 

An intelligent strategy for an attacker is to launch a series of selfish mining attacks and, once successful, combine them with a double-spending attack. This can be done by conducting regular public transactions while always having conflicting items hidden in the attacker's secret blocks. It is always possible that the recipient will accept the payment until a successful selfish mining attack is completed. In this case, in addition to the selfish mining, the double spending attack will also be successful. Thus, having a miner that for him a selfish mining strategy is at least as profitable as honest mining generally undermines the security of Bitcoin payments \cite{R4}, as the attacker pays no cost to attempt double-spending attacks and will eventually succeed in shaping the attack. On the other hand, an attacker who cannot benefit from selfish mining alone may find it profitable to combine this strategy with double-spending attacks, which has potentially serious consequences for the selfish mining profitability threshold.

This paper proposes a combined attack model and two models to defend against this attack. Our methods are based on a new fork-resolving policy that uses a new weighting algorithm to choose the winning chain in forks.

Our defense methods replace the original Bitcoin fork-resolving policy, represented by length FRP, with smart FRP. At any moment, according to the conditions of the network, miners choose one of two weight or length criteria to determine the winning chain in a fork. In weighted FRP, the miner assigns a weight to the blocks in each chain according to the time stamps, and the chain with the most weight is selected as the main chain. On the other hand, we will see that the weight criterion will only sometimes be the best choice, and conditions must be created to use this criterion only in exceptional cases. Furthermore, the algorithms presented in this paper intelligently set a parameter known as the number of confirmations required by the service provider to send the goods in order to reduce the risk of double-spending attacks. 

Compared with existing defenses, our defense methods have many advantages. Our methods are backward-compatible, decentralized, effective, and can reduce the probability of choosing a winning chain under the influence of eclipse attacks. For evaluation, we extend the SM1 developed by Eyal et al. \cite{R3} Results show that our methods successfully reduce the risk of SM1 and double-spending attacks.

\section{Preliminaries}
In this section, we first describe some definitions, then mention the defenses proposed before.
\subsection{Bitcoin Blockchain and Mining}
To better understand the presented algorithms, in this section we summarize the basic features of Bitcoin by reviewing the original paper by Nakamoto \cite{R2} and the book by Narayanan et al. \cite{R5} to have a complete view of the system. All nodes follow the same block and transaction validation rules to ensure participants' consensus on valid transactions. A typical Bitcoin transaction consists of at least one input and one output. The transaction fee, which is the difference between the total amount of inputs and outputs in a transaction, is paid to the miner, who records the transaction in the blockchain. 
\subsection{Selfish Mining}
In this attack, a set of miners use dishonesty in the consensus process and conspire with each other to inject a set of decisions and fake blocks into the network.

The scenario of this attack is that when honest miners discover a new block, (1) if the size of the public chain (the honest branch) is longer than the selfish branch (the private chain created by the attacker), then selfish miners try to set their private branch to the public branch. (2) If the selfish branch is one block longer than the public branch, then selfish miners will fully publish their private chain. (3) If the selfish branch is more than one block longer than the public branch, then selfish miners only publish the first block of their private branch. When selfish miners discover a new block, they keep it private, and when competing with honest miners, they publish their private branch to win the race \cite{R3}.
\subsection{Double-Spending attack}
Bigam et al. \cite{R6} have described a double-spending attack in five steps. A double-spending attack refers to the fact that a different number of transactions have occurred where the cryptocurrencies are the same. It means that one unit of digital currency has been spent twice in two different transactions. The five steps for a double-spending attack that Begam et al. described are as follows.
\begin{enumerate}
    \item The process of adding blocks. First, the user requests a transaction through his wallet. This unconfirmed transaction is placed in a pool of unconfirmed transactions. We know that miners select transactions and add blocks to the blockchain by solving complex mathematical proof-of-work problems. So this block will be added only if other miners confirm the obtained hashes.

    \item Once the honest miners have verified the block and the block has been added to the main blockchain, the attacker's miner starts his chain with the verified block. This miner spends all his currency and sends this information to the main blockchain but does not put it on his private chain.
    \item In this step, the attacker selects transactions and adds the block to his private chain by verifying them with his powerful computing power faster than honest miners can add the block to the real blockchain.
    \item When the attacker broadcasts his private chain transactions in the main blockchain, if the private chain is larger than the real chain, honest miners on the real chain will try to add their block to the newly discovered chain as well.
    \item The rule governing the blockchain states that blocks are added to the head block of the larger chain by removing previous records. Since the head block of the real blockchain has information about the transaction in which the corrupt miner spent his currency, the private chain does not know about the first transaction. Therefore, the previous transaction information is deleted when the private chain wins. Thus, in the new private chain, the attacker can re-spend all the coins he spent once in the real blockchain.
\end{enumerate}

According to the proposed solutions to deal with double-spending attacks, today, before sending the service, the service providers wait until the confirmation of a certain number of blocks to reduce the possibility of invalidating the confirmed blocks, and then proceed to send the service. This threshold limit is considered equal to 6 in Bitcoin, so to create such an attack, the attacker's private chain needs to be larger than the actual chain, in addition to being larger than this threshold value, so that the multi-confirmation double-spending attack occurs.
\subsection{Combined attack} As it is known, the double-spending attack alone requires very high computing power and cost, but by combining it with selfish mining, this attack can be carried out with far less computing power \cite{R4,R21}. In order to check our proposed solution to deal with these attacks, we first need to have a proper simulation of these attacks. In this paper, it is suggested that in selfish mining, if the attacker's chain surpasses the chain of honest miners and the length of the public chain is more than the number of confirmations needed for the merchant to send goods, it is a strategic moment for executing a double-spending attack. This is due to the possibility that the merchant has already sent the goods to the attacker after receiving the necessary confirmation.
In our defensive strategies, we try to avoid such situations as much as possible. Algorithm \autoref{CA} is the pseudocode of this section. \autoref{tblCA} shows the notations used in all of the pseudocodes used in this paper.
\begin{table}[htbp]
  \centering
  \caption{Notations of Pseudocodes.}\label{tblCA}
  \begin{tabularx}{0.9\linewidth}{XX}
    \toprule
    Name & Denotes \\
    \midrule
    PBL & Private Branch Length  \\
    DS & Number of opportunities to perform a DS attack  \\
    NRC & \#Required confirmations before sending the goods\\
    LA\_K & SM-related Learning Automata  \\
    LA\_Z  & DS-related Learning Automata  \\
    K &  SM-safe parameter \\
    Z & DS-safe parameter \\
    Beta\_K & Reinforcement signal of SM-related LA \\
    Beta\_Z & Reinforcement signal of DS-related LA \\
    N & Number of chains in a fork  \\
    Ch[$i$] & $i^{th}$ element of chains array in $\tau$  \\
    CW[$i$] &  $i^{th}$ element of chains weight \\
    CL[$i$] &  $i^{th}$ element of chains length \\
    $L_M$ &  Max length of forks in the chains array \\
    MTI & Max TimeStamp Index \\
    TS & TimeStamp \\
    $C_H$ &  last block's height of the main chain before the fork \\
    LA & Learning Automata  \\
    L & Action selected by LA \\
    $Z_{\max}$ & Max value for the DS-safe parameter  \\
    $Z_{\min}$ & Min value for the DS-safe parameter \\
    SBCR & Stale Blocks Change Rate\\
    CVW[$i$] &  $i^{th}$ element of chainsvalidatingweight \\
    Threshold & Threshold of a valid chain \\
    Beta & Reinforcement signal \\
    \bottomrule
  \end{tabularx}
\end{table}
\begin{algorithm}
  \caption{Combined Attack} \label{CA}
  \begin{algorithmic}[1]
    \Initialize{PublicChain $\gets$ Publicly Known Blocks \\  PrivateChain $\gets$ Publicly Known Blocks
    \\  PBL $\gets$ 0
    \\  DS $\gets$ 0
    \\  Mine at the head of the private chain}
    \State Attacker sends a transaction to the vendor
    \If{(SelfishPool found a block)}
        \State \small $\Delta_{prev} \gets$ length(PublicChain) - length(PrivateChain)
        \State Append new block to PrivateChain
        \State PBL $\gets$ PBL + 1
        \If{$\Delta_{prev}$=0 \: and  \: PBL=2}
            \State /*Was tie with branch of 1*/
            \State Publish all of the PrivateChain
            \State PBL $\gets$ 0
        \EndIf
        \State Mine at the new head of the PrivateChain
    \ElsIf{(Other miners found a block)}
        \State \small $\Delta_{prev} \gets$ length(PublicChain) - length(PrivateChain)
        \State Append new block to PublicChain
        \If{$\Delta_{prev}$=0}
            \State PrivateChain $\gets$ PublicChain
            \State PBL $\gets$ 0
        \ElsIf{$\Delta_{prev}$=1}
            \State Publish the last block of the PrivateChain
        \ElsIf{$\Delta_{prev}$=2}
            \State Publish all of the PrivateChain
            \If{length(PulicChain) $>=$ NRC}
                \State DS $\gets$ DS+1
            \EndIf
        \Else
            \State Publish first unpublished block in private block
        \EndIf
    \State Mine at the head of the PrivateChain
    \EndIf
    \end{algorithmic}
\end{algorithm}
\subsection{Learning Automata} 
\begin{figure}
	\centering
		\includegraphics[scale=.75]{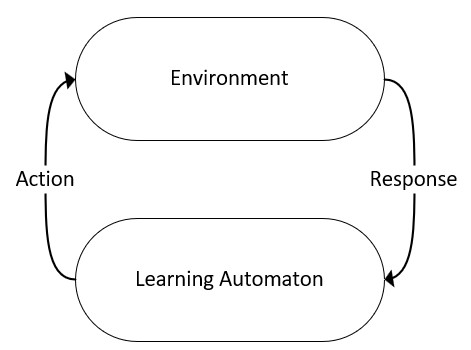}
	\caption{Interaction between the LA and the environment.}
	\label{LAFig}
\end{figure}

A learning automaton(LA)\cite{R7} is a machine that can perform a finite number of operations. A possible environment evaluates every action chosen, and the result of this evaluation is given to the learning automaton in the form of a positive or negative signal, and the automaton is affected by this response to choose the next action. The ultimate aim of the process is for the automaton to learn how to select the optimal action from its available options, by maximizing the probability of obtaining rewards from the environment. In this paper, we use two types of Variable Depth Hybrid Learning Automaton (VDHLA) proposed by Nikhalat-Jahromi et al \cite{R8}. \autoref{LAFig} illustrates the interaction between the learning automaton and the environment.
\subsection{Properties of an Ideal Defense} We can enumerate desirable properties of an ideal defense by explaining the problems and weaknesses of existing defenses \cite{R9}.
\begin{enumerate}[\textbullet]
\item Decentralization: The introduction of a trusted server would lead to a new single point of failure, which would be contrary to Bitcoin’s fundamental principles.
\item Incentive Compatibility: The expected relative revenue of a miner should be proportional to his mining power.
\item Backward compatibility: Individuals who do not engage in mining activities and are unable to upgrade their clients are still able to take part in the network, which is particularly vital for hardware products like Bitcoin ATMs. It is crucial that the following regulations remain unchanged.
\begin{enumerate}
    \item Block validity rules: A block that meets the requirements of the current Bitcoin protocol must also meet the requirement of the new method.
    \item Reward distribution policy: All blocks in the main chain and no other block receive block rewards.
    \item Eventual consensus: Despite potential attacks, both new and old clients are ultimately expected to reach a consensus on the main chain.
\end{enumerate}
\end{enumerate}

\subsection{Existing Defenses}
To prevent a combined attack, selfish mining must be prevented. To do this, we will first talk about the ways these attacks can be prevented.
\subsubsection{Uniform tie-breaking}
Eyal and Sirer proposed a defense mechanism for resolving ties in the mining process. This involves miners randomly selecting which chain to mine on in the event of a tie. The authors demonstrated in their research that this approach effectively increases the profit threshold for earning unfair block rewards within the selfish mining strategy to $25\%$ \cite{R3}. The defense is also referred to as uniform tie-breaking \cite{R4}. There are two notable drawbacks to this particular defense mechanism. Firstly, it can aid attackers with an advertisement factor of less than $0.5$, which is the fraction of computing power from honest nodes that would accept a block from a selfish miner over one from an honest miner (represented by parameter $\gamma$). Secondly, even with a profitable threshold of $25\%$, this approach remains risky in the context of small blockchains.
\subsubsection{Ethan Heilman’s proposed method}
Heilman proposed a method based on unforgeable timestamps against selfish mining called "Preferred Freshness." This method requires miners to add unforgeable timestamps to blocks, encourages honest miners to choose blocks with the latest timestamp, and invalidates blocks hidden by attackers. However, the disadvantage of this method is that it requires a valid timestamp agent to generate unforgeable timestamps, which requires honest miners to log all last timestamp release reports \cite{R10}.
\subsubsection{Publish or Perish}
Zhang and Preneel proposed a weighted fork-resolving policy. When a fork occurs, a weight is calculated for each chain, and it is recommended that honest miners not rely on chain length when choosing the main chain but choose the chain with the highest weight \cite{R9}. Below are listed the disadvantages of this method.
\begin{itemize}
    \item The excessive emphasis is placed on the weight criterion for selecting the winning chain in a fork.
    \item This method relies on a fixed value for parameter $K$, which plays a crucial role.
    \item The potential for developing novel forms of attacks.
    \item Despite the paper's assertion that there has been no change in the structure, it is imperative to make significant alterations to both the blocks' structure and the consensus protocol.
\end{itemize}
\subsubsection{Saad’s proposed method}
By measuring transaction size, transaction fees, and other factors, Saad et al. assigned an "expected confirmation height" (i.e., the expected height of the block containing a given transaction) to each transaction. The smaller the gap between the actual confirmation and expected height, the less likely the selfish mining behavior is \cite{R11}. Some modifications to the Bitcoin block structure and transactions are required for this method, which may become its primary drawback.
\subsubsection{Lee’s proposed method}
By adding transaction creation time to the transaction data structure, Lee and Kim increased the total hash required for the selfish mining profit threshold from $25\%$ to $33\%$ \cite{R12}. This approach may have some drawbacks, such as requiring modifications to the transaction data structure, which could be a significant disadvantage. Additionally, the profit threshold equal to $33\%$ may still pose a risk for smaller blockchains.
\subsubsection{Nik Defense} Nikhalat-Jahromi et al. proposed a time stamp-based weighted fork-resolving policy. When a fork occurs, a weight is calculated for each chain and based on a safe parameter calculated by learning automata; honest miners choose the main chain based on the chain’s length or weight \cite{R13}.
Below are listed the disadvantages of this method.
\begin{itemize}
    \item The excessive emphasis is placed on the weight criterion for selecting the winning chain in a fork.
    \item The inadequate definition of the reinforcement signal, coupled with the failure to account for sudden changes in the attacker's hash rate, can lead to an inappropriate setting of the $K$ parameter.
    \item The potential for developing novel forms of attacks.
    \item The weight criterion contains calculations that are not necessary.
\end{itemize}

There are also some studies that try to identify factors in order to detect selfish mining attacks \cite{R14,R15}. These papers use existing data of selfish mining attacks to create training and test data. But because of the stochastic nature of blockchain, it seems that we cannot be sure about the validity of these kinds of datasets.

There are several studies that can be used to prevent double-spending attacks on the blockchain. For example, we can refer to the papers by Ghassan et al. \cite{R16,R17} and the paper by Podolanko et al. \cite{R18} to prevent double-spending attacks in fast payments. In these papers, solutions such as using enhanced observers and nearby peers to detect and warn about double-spending attacks are proposed. Bamert et al. \cite{R19} have also proposed two countermeasures against double-spending attacks. For example, the merchant must connect to a sufficient number of random nodes in the Bitcoin network. This makes it difficult for an attacker to inject false transaction information or a transaction containing a double-spending attack because the attacker does not know which nodes the merchant is connected to. Additionally, the merchant should not accept direct input connections. Therefore, the attacker cannot directly send a transaction to the merchant and is forced to broadcast it over the network. Nodes transmitting that transaction will check and detect if there is a double-spending attempt. Subsequent transactions from the attacker using the same input (address) are ignored by those nodes, making it difficult for the attacker to perform a double-spending attack.

The main solution to prevent double-spending attacks is waiting for multiple confirmations. In this method, the vendor should wait for multiple confirmations before releasing the product or providing a service to the client \cite{R20}. The idea of waiting for more confirmations to prevent double-spending attacks has been discussed in many papers; this value is set to 6 for Bitcoin traders by contract. Arthur Gervais et al. \cite{R21} have investigated the changes of $v_d$, which is the lowest value of a double-spending transaction for the profitability of a combined attack, with different values of this parameter. 

It is also mentioned in another paper \cite{R22} that it is possible to change the mentioned parameter in order to increase security and reduce the risk of double spending in the blockchain. The paper states that one way to counter these attacks is to allow servers to require more confirmations for larger transactions and make the attack more difficult.

\section{Proposed Algorithms}

The proposed algorithms are presented in this section. We begin by describing how combined attacks and the proposed defense work. The proposed algorithms are then explained with the required definitions. Lastly, we demonstrate the proposed algorithms.
\subsection{Main Idea}
As previously discussed, the combination of double-spending and selfish mining attacks poses a significant threat. Therefore, an effective defense system must be capable of detecting such attacks. Various studies, including Gervis et al.'s experiments \cite{R21}, have established a direct correlation between selfish mining and the rate of stale blocks. When an attacker is present in the network, the rate of stale blocks increases as a result of forks being won by one chain over another. This heightened rate of stale blocks increases the risk of double-spending attacks, as confirmed by \cite{R21}. By observing and analyzing the rate of stale blocks generated by forks, we can gauge the vulnerability of the network and use this data to train our model. It is important to focus on stale blocks generated by forks for two reasons: 
\begin{enumerate}
    \item We cannot know about stale blocks that the attacker chooses not to publish.
    \item Stale blocks resulting from an attacker's surrender without competition do not necessarily indicate danger to the network. Such attacks only pose a threat if the attacker has infiltrated an honest pool.
\end{enumerate}
In this paper, we propose the utilization of the stale block rate to train our intelligent model as our primary idea.

\subsection{System Model}

We propose a model that prioritizes miner nodes as the key players in the blockchain network while disregarding other nodes such as super nodes, light nodes, and others. In this model, a selfish mining attack occurs when mining nodes unite to form a mining pool with the goal of obtaining more revenue than they are entitled to. To create the optimal and most effective model, we consider two groups of miners: those who follow a selfish mining strategy, possessing less than 50\% of the total computing power, and those who adhere strictly to the Bitcoin mining protocol. We explain our model by expanding the one described in Nikhalat-Jahromi et al. \cite{R13}.

To describe how computing power is distributed in the proposed model, we assume that the selfish mining pool holds a fraction $\alpha$ of the total computing power, while the remaining honest miners hold a fraction of $1$ - $\alpha$. Consequently, the probability that a newly discovered block belongs to the selfish pool is $\alpha$, and the probability that it belongs to other miners is $1$ - $\alpha$.

In our proposed model, we have made the assumption of disregarding the block propagation delay in order to improve the clarity of network connectivity. This assumption is justifiable as Bitcoin miners strive to transmit and receive blocks promptly, as any delay in this process can interrupt their mining schedule and affect their efficiency in finding new blocks. Moreover, there are ongoing efforts by researchers and network developers to minimize the propagation delay in the Bitcoin network, as evidenced by various published articles and improvement proposals \cite{R9}. In the subsequent paragraph, we will delve into the topic of block order in the network.

The blocks in a Bitcoin node are organized in a tree structure, with each block containing a reference to the previous one. To simplify our analysis, we will focus on just two branches of the block tree: the main chain, which is the longest chain agreed upon by consensus among nodes, and a private branch created by selfish miners. It is impossible for an honest node to differentiate between these two branches. The following paragraphs will detail the miner behavior within this network model. 

In decentralized networks such as Bitcoin that operate on a proof-of-work consensus mechanism, the creation of a new block can be regarded as an event that is not influenced by the passage of time. As a result, the mining process is considered both discrete and memoryless. This implies that at the moment of discovering a new block, every miner, regardless of their honesty or selfishness, makes a decision that persists until the next block is found.

Our proposed model involves the selfish miner utilizing their computing power to create a private chain. At a given time t, the selfish miner must decide which block from the main chain to extend their private chain with and which block to release in order to increase the selfish pool's revenue. 

If the selfish miner becomes aware that an honest miner has discovered a new block, they may attempt to substitute their private block. The advertisement factor, represented by parameter $\gamma$, is a crucial element of this model. $\gamma$ is the fraction of computing power belonging to honest nodes that would accept a block from a selfish miner instead of an honest miner. When considering the block's height within Bitcoin's network ($h$), the probability of the selfish miner's block being accepted by honest nodes at that height is $\gamma$\*($1$-$\alpha$\*). Each miner node within the network utilizes one or two learning automata. The following sections will detail the proposed algorithm's definitions based on the $i^{th}$ miner in the network.
\subsection{Required Definitions}
In this section, the required definitions of the algorithm are defined in order to explain the proposed algorithm.
\begin{itemize}
    \item \textbf{Decision-making time ($\tau$):} This definition outlines the concept of fork decision-making time from the perspective of the $i^{th}$ miner. This involves a designated period for the miner to assess whether any forks exist and, if so, to choose between them. To specify this time parameter in the proposed method, it is referred to as $\tau$. 
    \item \textbf{Time Window:} The interval during which the miner chooses to adjust the safe parameters K and Z is referred to as the time window. This window is determined by an integer factor of the time parameter $\tau$.
    \item \textbf{SM safe parameter ($K$):} If the difference in length between the two longest chains in the fork is less than $K$, the chain with the higher weight wins in the fork. 
    \item \textbf{DS safe parameter ($Z$):} In the proposed methods, the parameter $Z$ denotes the value that merchants are recommended to obtain confirmation for prior to dispatching goods, to mitigate the risk of double spending.
    \item \textbf{Timestamp:} Current time as seconds in the universal time since January 1, 1970. Each block contains a timestamp whose main function is to determine the exact moment in which the block has been mined and validated by the blockchain network.
\end{itemize}
In the subsequent sections, we will present the proposed methods based on the system model and the definitions outlined.

\subsection{Smart defense system with two learning automata (SDTLA)}
This section will provide a detailed overview of the first proposed algorithm. The algorithm will be initially explained through a sequence of events that occurred during the defense process, followed by a discussion of sub-algorithms. The proposed algorithm is capable of responding to the following events:
\begin{enumerate}
    \item One Block Receive Event 
    \begin{enumerate}
        \item If a fork exists, a new block's relation to existing forks will be checked by the previous hash parameter. Forks will be created if necessary by the miner.
    \end{enumerate}
    \item Decision-Making Time ($\tau$) Event 
    \begin{enumerate}
        \item The existence of a fork will be checked, and if present, it is essential to determine the selection criteria, either by length or weight.
    \end{enumerate}
    \item Time Window Event
    \begin{enumerate}
        \item The existence of a fork will be checked, and if present, it is essential to determine the selection criteria, either by length or weight.
        \item The reinforcement signals will be used to update the learning automatons.
        \item The next actions will be determined by the learning automatons, and the safe parameters will be updated accordingly.
    \end{enumerate}
\end{enumerate}
The proposed algorithm consists of five sub-algorithms. These five sub-algorithms will describe in the following.
Algorithm \autoref{SDTLA} is the pseudocode of this section. 

\begin{algorithm}
  \caption{SDTLA}\label{SDTLA}
  \begin{algorithmic}[1]
   \Switch{(Event)}
        \Case{‘TimeWindowEvent’}
        \State CalculateStaleBlocksRateperZ\*(\:\*)
        \State CalculateStaleBlocksRateperK\*(\:\*)
        \State ForkCreationChecking\*(\:\*)
        \State Beta\_K$\gets $ CalculateSMUpdateSignal\*(\:\*)
        \State Beta\_Z$\gets $ CalculateDSUpdateSignal\*(\:\*)
        \State LA\_K.update\*(Beta\_K)
        \State LA\_Z.update\*(Beta\_Z)
        \State UpdateSmSafeParameter\*($K_{min}, K_{max}$\*)  
        \State UpdateDsSafeParameter\*($Z_{min}, Z_{max}$\*)
        \EndCase
        \Case{‘TauEvent’}
        \State ForkCreationChecking\*(\:\*)
        \EndCase
        \Case{‘BlockReceiveEvent’}
        \State /*Just put block on the correct fork’s chain and If needed, use the ForkSelection algorithm()*/
        \EndCase
        \EndSwitch
   \State \textbf{Display} ($Z$)
  \end{algorithmic}
\end{algorithm}

\subsubsection{Length Calculation}
Selfish mining creates a fork, as we all know. The length of every chain created by the fork condition is one characteristic that can be used to defend against it. For calculating the length of each chain created by the fork, first get the height of the last block before the fork was created, then calculate the difference between that height and the height of the last block before the fork \cite{R13}.
\subsubsection{Weight Calculation}
To defend against selfish mining, the weight of every chain created by the fork condition can also be used. In this paper, we propose calculating the weights by considering the first ten blocks of each chain in the fork and assigning greater weight to older blocks. The rationale behind this approach is that the attacker's selfish behavior becomes evident in the initial blocks, which were previously concealed. For each chain created by forking, the following steps will be taken: 
\begin{enumerate}
    \item The first ten blocks (the oldest) are selected from each chain.
    \item Based on the maximum length, evaluate blocks of different chains but of the same height. The chain with the most recent timestamp will win the race; so, among the blocks present at the same height, the weight of the chain that has the highest time stamp, according to the height at which it is located, is added to the value that is the result of multiplying one unit by the corresponding coefficient of that height (because the lower timestamp means that the block is older and more likely to be hidden by the attacker, and also the heights associated with older blocks should have a higher coefficient because they have a higher value in weighting).
    \item The calculation in part 2 will continue until the maximum height or the tenth block has been reached; we consider ten blocks for weighting. A shorter chain will not conclude in comparisons of blocks with higher heights if the others are longer. In the end, if the length of the chains is more than ten, taking into account that, in this case, the superiority of the length of the chain is not considered in the weight, it is necessary to take action to solve this problem. Therefore, the amount of difference is calculated from ten and is added to the weight of the chain by multiplying by 0.5 (a value less than the lowest weight of the first ten blocks).   
\end{enumerate}
Algorithm \autoref{WC} is the pseudocode of this section. 
\begin{algorithm}
  \caption{Weight Calculation} \label{WC}
  \begin{algorithmic}[1]
      \For{$i \gets 1$ to min\*($L_M,10$\*)}
        \State $MTI\gets 1$
        \For{$j \gets 1$ to $N$}
            \If{$Ch[j][i].TS > Ch[MTI][i].TS$}
                \State $MTI\gets j$
            \EndIf
            \State \small W = [1, 0.95, 0.9, 0.85, 0.8, 0.75, 0.7, 0.65, 0.6, 0.55]
            \State CW[MTI] $\gets$ CW[MTI]+\*($1 * W[i]$\*)
        \EndFor
      \EndFor
      \For{$j \gets 1$ to $N$}
        \If{$CL[j] > 10$}
            \State CW[j]$\gets $ CW[j] + \*($CL[j]$-$10$\*) $*$ 0.5 
        \EndIf
    \EndFor
    \State \Return CW
\end{algorithmic}
\end{algorithm}
\subsubsection{Chain Selection}
The miner must make a decision among the chains created by the fork condition. Our first method employs the chain selection approach proposed by Nikhalat-Jahromi et al. \cite{R13}. The chain selection algorithm of the proposed defense can be described as follows:
\begin{enumerate}
    \item Calculating the chain length.
    \item Sorting chains based on length in descending order.
    \item If one chain is longer than the others by $K$; it will select it for the next mining event.
    \item If no chain is longer than the others by $K$, the weight of all chains will be calculated by the algorithm described before.
    \item Sorting chain based on the weight in descending order 
    \item Heaviest chain will be selected.
\end{enumerate}
Algorithm \autoref{CSSDTLA} is the pseudocode of this section. 
\begin{algorithm}
  \caption{Chain Selection} \label{CSSDTLA}
  \begin{algorithmic}[1]
    \If{$N > 1$}
        \State CL$\gets$ ForkChainsLengthCalculation\*($C_H$\*)
        \State SortDescendingly\*($Ch, CL$\*)
        \If{CL[0] - CL[1] $> K$}
            \State /*Decide based on Chain’s Length*/
            \State ChosenChain $\gets$ Ch[0]
            \State $C_H \gets$ ChosenChain.LastBlockHeight
        \Else
            \State /*Decide based on Chain’s Weight*/
            \State $L_M \gets$ CL[0]
            \State CW $\gets$ ChainsWeightCalculation\*($L_M$\*)
            \State SortDescendingly\*($Ch, CW$\*)
            \State ChosenChain $\gets$ Ch[0]
            \State $C_H \gets$ ChosenChain.LastBlockHeight
        \EndIf
        \State \Return ChosenChain
    \EndIf
    \end{algorithmic}
\end{algorithm}
\subsubsection{Action selection by LA}
With the end of the time window, considering that the time window is an integer coefficient of the time interval $\tau$ and in each time interval $\tau$, unique events have happened, it is necessary to make a decision about the outcome of these events. Decision-making, in this case, is the responsibility of learning automata. The learning automatons at the end of the time window, according to the selected action, adjust the value of safe parameters according to the determined interval. In this method, we use two learning automata.
\begin{enumerate}
    \item Action selection by SM-LA: In order to update the SM safe parameter, we use a learning automaton. This learning automaton is responsible for setting the K parameter. The action that the learning automaton chooses is one of the following three actions:
    \begin{enumerate}
        \item Grow: This action happens if the network is considered to be under a selfish mining attack, which means that the rate of change of stale blocks has increased compared to $K$ and should be fixed. In this case, the $K$ parameter increases by one unit.
        \item Stop: This action happens when the learning automaton concludes, according to the received signal, that the value of the safe parameter related to selfish mining was appropriate in the previous time period. So there will be no need to update $K$.
        \item Shrink: This action happens if the network is less exposed to selfish mining attacks, which means that the rate of change of stale blocks decreases compared to $K$. In this case, the $K$ parameter is reduced by one unit.
    \end{enumerate}

Algorithm \autoref{SMLA} is the pseudocode of this section. 
\begin{algorithm}
  \caption{UpdateSmSafeParameter\*($K_{min}, K_{max}$\*)} \label{SMLA}
  \begin{algorithmic}[1]
    \If{$K = K_{max}$}
        \State L$\gets$ LA.action\*([‘Stop’, ‘Shrink’]\*)
    \ElsIf{$K = K_{min}$}
        \State L$\gets$ LA.action\*([‘Grow’, ‘Stop’]\*)
    \Else
        \State L$\gets$ LA.action\*([‘Grow’, ‘Stop’, ‘Shrink’]\*)
    \EndIf
    
    \Switch{(L)}
        \Case{‘Grow’}
            \State K $\gets $ K + 1
        \EndCase
        \Case{‘Shrink’}
            \State K $\gets $ K - 1
        \EndCase
        \Case{‘Stop’}
            \State /*Do nothing about K*/
        \EndCase
    \EndSwitch
    \end{algorithmic}
\end{algorithm}
    \item Action selection by DS-LA
    In order to update the DS safe parameter, we use a learning automaton. This learning automaton is responsible for setting the $Z$ parameter. The action that the learning automaton chooses is one of the following three actions:
    \begin{enumerate}
        \item Increase: This action happens if the network is under attack by selfish mining and double-spending attackers, which means that the rate of changes of stale blocks has increased compared to $Z$ and should be fixed. In this case, determining the $Z$ parameter has two modes.
        \begin{itemize}
            \item If the Stale Blocks Change Rate is greater than 0.75, it indicates that the network is heavily exposed to attack and the need to quickly reduce the rate of change is felt, so we multiply the value of Z by 2 to have a sudden jump.
            \item Otherwise, even though the rate of change is increasing and the need to reduce it is felt, there is no need for drastic changes, and it is enough to increase the Z parameter by 2 units.
        \end{itemize}
        \item No Change: This action happens when the learning automaton concludes, according to the received signal, that the value of the safe parameter related to the double-spending attack in the previous time period was appropriate. So there will be no need to update $Z$.
        \item Decrease: This action happens if the network is less under the attack of selfish mining and double spending, which means that the rate of changes of stale blocks has a downward trend compared to $Z$ and the speed of transaction confirmation can be increased by reducing $Z$. In this case, determining the $Z$ parameter has two modes.
        \begin{itemize}
            \item If $Z$ is greater than 6 and divisible by 2, we divide this parameter by 2.
            \item Otherwise, we subtract 2 units from it.
        \end{itemize}
        The reason for this way of dealing with the $Z$ parameter is that the fact that this parameter is divisible by 2 can be a sign that it has had an instantaneous jump before (multiplied by 2 in the Increase operation) and has not experienced a sharp decrease in dividing by 2.
    \end{enumerate}
\end{enumerate}
In this section, we use the term "Stale Blocks Change Rate," which is defined based on the current Z in the given time window. \autoref{eq:2} shows the calculation of this term. 
\begin{equation}
    \begin{aligned}
        \label{eq:1}
            \scalebox{0.93}{$StaleBlockRatePerZ=\frac{\text{\#}fork Stale Blocks In Window}{Current Z}$}
    \end{aligned}
\end{equation}

\begin{equation}
    \begin{aligned}
        \label{eq:2}
            \scalebox{0.94}{$SBCR=\frac{StaleBlockRatePerZ_{New}}{StaleBlockRatePerZ_{New} + StaleBlockRatePerZ_{Old}}$}
    \end{aligned}
\end{equation}
\\
\autoref{eq:1}, defined here, is also utilized for computing the reinforcement signal in double-spending related learning automata, as detailed in the subsequent sections.
Algorithm \autoref{DSLA} is the pseudocode of this section. 
\begin{algorithm}
  \caption{UpdateDSSafeParameter\*($Z_{min}, Z_{max}$\*)} \label{DSLA}
  \begin{algorithmic}[1]
    \If{$Z >= Z_{max}$}
        \State L$\gets$ LA.action\*([‘No Change’, ‘Decrease’]\*)
    \ElsIf{$Z <= Z_{min}$}
        \State L$\gets$ LA.action\*([‘Increase’, ‘No Change’]\*)
    \Else
        \State \small {L$\gets$ LA.action\*([‘Increase’, ‘No Change’, ‘Decrease’]\*)}
    \EndIf
    \Switch{(L)}
    \Case{‘Increase’}
        \If{$SBCR >= 0.75$ and $Z <= Z_{max}$}
           \State Z$\gets$ Z * 2
        \Else
            \State Z$\gets$ Z + 2
        \EndIf
        \EndCase
    \Case{'Decrease'}
        \If{$Z > 6$ and $Z$\%$2 = 0$}
           \State Z$\gets$ Z / 2
        \ElsIf{$Z > Z_{min}$}
            \State Z$\gets$ Z - 2
        \EndIf
        \EndCase
    \Case{'No Change'}
        \State /*Do nothing about Z*/
    \EndCase
    \EndSwitch
    \end{algorithmic}
\end{algorithm}

\subsubsection{Calculate Update Signals}
As we said before, we have to update the signals. One for the SM learning automaton and another one for the DS learning automaton. This section defines the calculation of these signals from the $i^{th}$ miners' point of view. The learning automata must calculate the reinforcement signal. In this paper, the signal is denoted by $\beta$. The reinforcement signal is feedback from the environment that the learning automaton uses to update its probability vector. In the proposed method, this signal is calculated from the analysis of each decision $\tau$ in a time window. 
\begin{enumerate}
    \item Calculate Update Signals for SM learning Automata: At the end of the time window, since the learning automaton has chosen its action according to the previous window, before selecting the next action, it needs to receive its reward if it chose correctly and be penalized if it chose incorrectly. For this purpose, the parameter for updating the reinforcement signal is used in the learning automaton. In this section, the reinforcement signal for the learning automaton related to selfish mining is defined for the first algorithm.
\begin{equation}
    \begin{aligned}
        \label{SM:1}
            \scalebox{0.92}{$StaleBlockRatePerK=\frac{\text{\#}fork Stale Blocks In Window}{Current K}$}
    \end{aligned}
\end{equation}
\begin{equation}
    \begin{aligned}
        \label{SM:2}
            \scalebox{0.88}{$\beta_1=\frac{Number of Weight Decision}{Number of Height Decision + Number of Weight Decision}$}
    \end{aligned}
\end{equation}
\begin{equation}
    \begin{aligned}
        \label{SM:3}
            \scalebox{0.95}{$\beta_2=\frac{StaleBlockRatePerK_{Old}}{StaleBlockRatePerK_{New} + StaleBlockRatePerK_{Old}}$}
    \end{aligned}
\end{equation}
If the values of $\beta_1$ and $\beta_2$ are greater than $0.5$, they will be equal to one. Otherwise, they will be equal to zero. Now we can calculate $\beta$ for SM learning automata:
\begin{equation}
    \begin{aligned}
        \label{SM:4}
            \beta = \beta_1  and  \beta_2
    \end{aligned}
\end{equation}
Finally, a value of one for $\beta$ means that the selection of the automata is correct, and a reward should be given to it. Otherwise, a penalty will be given to the learning automata. As it is known, in this case, the automata will be rewarded only when the change rate of stale blocks has a downward trend compared to the K parameter. In this way, changes in the attacker's mining power will affect the model’s training.

    \item Calculate Update Signals for DS learning Automata: In this section, the reinforcement signal for the learning automaton related to the double-spending is defined for the first algorithm.

\begin{equation}
    \begin{aligned}
        \label{DS:1}
            \scalebox{0.95}{$\beta=\frac{StaleBlockRatePerZ_{Old}}{StaleBlockRatePerZ_{New} + StaleBlockRatePerZ_{Old}}$}
    \end{aligned}
\end{equation}
    If the $\beta$ is greater than $0.5$, the automaton selection is correct, and a reward should be given to it. Otherwise, a penalty will be given to the learning automata. As is known, in this case, the automaton will be rewarded only when the rate of change of stale blocks has a downward trend compared to the $Z$ parameter. In this way, changes in the attacker's mining power will affect the model’s training.
    \end{enumerate}
In the experiment section, we will see that this algorithm has several problems. To solve these problems, we introduce another model in the following.
\subsection{Weight validating based method (WVBM)}
In this section, we present a completely different solution that uses a new concept called weight threshold to deal with selfish mining attacks and somehow validates the chains. Unlike the previous system, this part of the system does not use learning automata and only uses learning automata to deal with double-spending attacks.
\subsubsection{Motivation}
The defense system presented in this section is designed to reduce calculations dependent on learning automata and get the results as close as possible to the ideal scenario, no individual would be able to earn revenue greater than their proportionate processing power. This would remove any incentive for rational miners to act selfishly. The two main features of this system are as follows:
\begin{enumerate}
    \item Using the weight threshold concept to validate the largest chain in a fork.
    \item Using just one learning automaton to calculate the $Z$ parameter and thus reduce double-spending attacks.
\end{enumerate}
The proposed weight threshold concept is independent of the weight calculated to select the winning chain, and the same innovative weighting proposed in the defense system of the previous model will be used for the weight. The defense system presented in this section must be able to provide all of the following:
\begin{itemize}
    \item Preserve the properties of the blockchain.
    \item Do not impose too much additional load on the nodes.
    \item Do not reduce the speed of transaction approval too much.
    \item The more nodes follow it, the more successful it will be.
    \item React correctly against the dynamic changes in the attacker's mining power.
    \item Train the $Z$ parameter correctly and reduce the risk of double-spending attacks.
    \item Reduce the success rate of selfish mining attacks as close as possible to the ideal scenario in which no individual would be able to earn revenue greater than their proportionate processing power by using the weight threshold concept.
    \item As much as possible, do not act randomly in selecting the winning chain in a fork
\end{itemize}
The most important thing about this system is that larger chains will win the fork if they have newer equivalent blocks than other chains in at least a certain percentage of their initial ten blocks. This prevents selfish miners who tried to hide their old blocks from winning the fork just because their private chain is bigger. If the largest chain does not have enough weight, the winner is selected based on the innovative weight introduced in the previous model. 
\subsubsection{The validating weight calculation}
To show the differences between this model and the former, we first present the validating weight. The validating weight calculation is given below.
\begin{enumerate}
    \item The first ten blocks (the oldest) are selected from each chain.
    \item Based on the maximum length, evaluate blocks of different chains but of the same height. The chain with the most recent timestamp will win the race. So, among the blocks present at the same height, the weight of the chain that has the highest time stamp will be added by one unit.
    \item Step 2 calculations continue for all ten heights considered in step one.
\end{enumerate}
Algorithm \autoref{VWC} is the pseudocode of this section. 
\begin{algorithm}
  \caption{Validating Weight Calculation} \label{VWC}
  \begin{algorithmic}[1]
      \For{$i \gets 1$ to min\*($L_M,10$\*)}
        \State MTI$\gets$ 1
        \For{$j \gets 1$ to $N$}
            \If{$Ch[j][i].TS > Ch[MTI][i].TS$}
                \State MTI $\gets$ j
            \EndIf
            \State CVW[MTI] $\gets$ CVW[MTI]+1
        \EndFor
      \EndFor
        \State \Return CVW
    \end{algorithmic}
\end{algorithm}
\subsubsection{Chain Selection}
This part is entirely different from the previous system. Chain selection in this algorithm includes the following steps:
\begin{enumerate}
    \item Calculate the length of the chains.
    \item Calculate the weight of the chains.
    \item The validation weight calculation.
    \item Sort chains by decreasing the length.
    \item If the validation weight calculated for the longest chain exceeds the threshold, that chain is selected. Otherwise, the chain with the highest weight is selected.
    \item If the lengths of chains are equal, the weight will be the criterion for choosing the winning chain.
\end{enumerate}
Algorithm \autoref{CSWVBM} is the pseudocode of this section.
\begin{algorithm}
  \caption{Chain Selection} \label{CSWVBM}
  \begin{algorithmic}[1]
    \If{$N > 1$}
        \State CL$\gets$ ForkChainsLengthCalculation\*($C_H$\*)
        \State SortDescendingly\*($Ch, CL$\*)
        \If{CL[0] $>$ CL[1] and CVW $>=$ Threshold}
            \State /*Decide based on Chain’s Length*/
            \State ChosenChain $\gets$ Ch[0]
            \State $C_H \gets$ ChosenChain.LastBlockHeight
        \Else
            \State /*Decide based on Chain’s Weight*/
            \State $L_M \gets$ CL[0]
            \State CW $\gets$ ChainsWeightCalculation\*($L_M$\*)
            \State SortDescendingly\*($Ch, CW$\*)
            \State ChosenChain $\gets$ Ch[0]
            \State $C_H \gets$ ChosenChain.LastBlockHeight
        \EndIf
        \State \Return ChosenChain
        \EndIf
    \end{algorithmic}
\end{algorithm}

In this system, we considered the weight threshold equal to a quarter of the length of the chain. In general, this threshold should be a function of the chain's length, so that it can be chosen based on the length of the chain and the level of risk that you pose to the network if the chain belongs to the attacker. A quarter of the length means that since the validation weight is an integer, if a chain has a length of 10, its minimum validation weight should be 3.\\
These two sections are the main differences between our new model and the former. 
Algorithm \autoref{WVBM} is the pseudocode of this section. 
\begin{algorithm}
  \caption{WVBM}\label{WVBM}
  \begin{algorithmic}[1]
   \Switch{(Event)}
        \Case{‘TimeWindowEvent’}
        \State CalculateStaleBlocksRateperZ\*(\:\*)
        \State ForkCreationChecking\*(\:\*)
        \State Beta$\gets $ CalculateDSUpdateSignal\*(\:\*)
        \State LA.update\*(Beta) 
        \State UpdateDsSafeParameter\*($Z_{min}, Z_{max}$\*)
        \EndCase
        \Case{‘TauEvent’}
        \State ForkCreationChecking\*(\:\*)
        \EndCase
        \Case{‘BlockReceiveEvent’}
        \State /*Just put block on the correct fork’s chain and If needed, use the ForkSelection algorithm()*/
        \EndCase
        \EndSwitch
   \State \textbf{Display} ($Z$)
  \end{algorithmic}
\end{algorithm}

\section{Evaluation}
In order to evaluate the efficiency of the presented algorithms, it is first necessary to introduce evaluation metrics. In the next section, experiments are designed, and by analyzing the results of these experiments according to the mentioned metrics, the quality of the proposed solution can be measured. We know that the attackers are unknown, and like other nodes, they have access to information about safe parameters, so to be fair, we consider the possibility that attackers release their private branches when the difference is higher than K. This kind of attack is proposed in \cite{R13} and is called Modified SM1. 

Taking inspiration from previous defense mechanisms and their simulators \cite{R3,R4,R9,R13}, we simulated the proposed algorithm by converting the mining model into a Monte Carlo simulation process. This conversion enables the distribution of newly discovered blocks among selfish and honest miners, without solving a cryptographic puzzle. The simulator generated 10,000 blocks for each experiment, testing varying selfish pool sizes. All the tests performed in this paper were performed on a computer equipped with an Intel Core i7 9750H processor with a working frequency of 2.6 GHz and using the Python programming language.
\subsection{Evaluation metrics}
In this section, several evaluation metrics are introduced to evaluate the intelligent defense model. These metrics will be used in the next section for experiments designed to validate the presented algorithms.
\subsubsection{Relative revenue of selfish miners}
Considering the attack by selfish miners, it is necessary to obtain the relative revenue of selfish miners. \autoref{ev:1} can be used to calculate the relative revenue of selfish miners.\\
\begin{equation}\label{ev:1}
\begin{gathered}
\scalebox{0.95}{$\frac{\text{\#}SelfishMinerWinBlocks}{\text{\#}HonestMinerWinBlocks +\text{\#}SelfishMinerWinBlocks}$ * 100}
\end{gathered}
\end{equation}
\subsubsection{The number of times a double-spending attack can occur}
A double-spending attack occurs when the length of the attacker's chain exceeds the length of the honest chain and the number of confirmations required before the merchant sends the goods. To calculate this metric, we put a counter in the code that increases by one every time the mentioned conditions are met. Finally, the calculated value in the network with the defense system and the network without the defense system will be a suitable measure to measure the power of our defense system. In order to properly check our defense systems, we have obtained this value in the presence of the Nik-defense system and compared it with our defense systems.
\subsubsection{The upper bound of the relative revenue of selfish miners}
If the attack is carried out under ideal conditions, \autoref{ev:2} is used to calculate the upper bound of the revenue of selfish miners.
\begin{equation}
    \begin{aligned}
        \label{ev:2}
            \frac{\alpha}{1 - \alpha}
    \end{aligned}
\end{equation}

\subsubsection{Profitable threshold}
The minimum ratio of hash power that brings more rewards to a selfish miner is called the profitable threshold.
\subsection{Experiments}
In this section, experiments have been designed and implemented to check the performance of the proposed models. The purpose of these tests is to check the performance of the models in the face of double-spending and selfish mining attacks. Each experiment consists of 1000 iterations, meaning that we consider 1000 mined blocks. Additionally, we repeat each experiment 50 times and report the average of the results as the final outcome to ensure a fair analysis. In our selfish mining experiments, we conducted a detailed analysis to accurately determine the profit threshold. We achieved this by examining the relative revenue of the selfish pool, considering various processing power values ($\alpha$) ranging from 0.20 to 0.5 with intervals of 0.02. These experiments were performed on networks that were fortified with our proposed defenses. In double-spending experiments, we conducted an analysis to determine the frequency of successful double-spending attacks. Specifically, we examined the behavior of a selfish pool with processing power $\alpha$ ranging from 0.20 to 0.45, with intervals of 0.05. These experiments were performed on networks equipped with our proposed defenses.
\subsubsection{The first experiment}
\begin{figure}
	\centering
		\includegraphics[scale=.6]{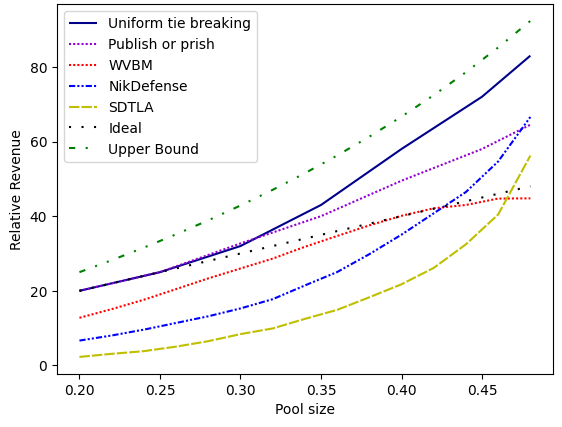}
	\caption{The first experiment, Compare our methods with previous works.}
	\label{FIG:1}
\end{figure}

This experiment aims to assess the effectiveness of the proposed models in protecting against selfish mining attacks. The results of this evaluation are presented in \autoref{FIG:1}, which illustrates the relative revenue earned by attackers based on their hash rates. The network is assumed to be equipped with the proposed models and compared to other defensive mechanisms, including the Nik-defense, Tie-breaking, and publish or perish systems.\\

In \autoref{FIG:1}, we can see the results of this experiment for both proposed methods. The black dotted line shows the ideal scenario. An ideal Scenario means a situation in which each miner is rewarded according to the amount of processing power they have shared. The blue line shows the Nik-Defense, so we can compare our methods to this method as the best method proposed before. In addition, we present the results obtained through the implementation of uniform tie-breaking and publish or perish techniques \cite{R9} to facilitate a more comprehensive analysis. As we can see in \autoref{FIG:1}, our first proposed method increases the profit threshold compared to previous works. But as it is clear from \autoref{FIG:1}, both the SDTLA presented by us and the Nik-Defense severely punish the attacker when they have even very little power. This indicates a relatively high-value selection for the safe parameter $K$ even when the network is not at serious risk. But this point can be hazardous. Because always choosing a weight criterion to choose your winner can open the way for new and even easier attacks compared to selfish mining. For instance, if an attacker persists in mining a block instead of abandoning their private chain, even when they lag two blocks behind, a high value for parameter $K$ and consequent selection of the winning chain based on weight criterion can lead to the attacker's triumph.

To address this issue, we present WVBM, which only applies weight criteria when necessary. Our weight validation-based approach predominantly utilizes length criteria for selecting the winning chain, ensuring that newer attacks cannot surpass selfish mining in terms of success rate. As depicted in \autoref{FIG:1}, the results of this method are comparable to those of the ideal scenario, establishing it as a robust defensive system against selfish mining. In an ideal scenario, no individual would be able to earn revenue greater than their proportionate processing power and as a result, there is no incentive for rational miners to act selfishly. In \autoref{tble1}, the profit threshold for each defensive mechanism is displayed. Our proposed methods, particularly WVBM, demonstrate the most favorable outcomes.
\begin{table}[h]
\centering
\caption{The profit threshold for each defensive mechanism.}
\label{tble1}
\begin{tabular}{cc}
\toprule
Model & Profit threshold \\
\midrule
Tie-breaking & 0.25 \\
Publish or perish & 0.25 \\
Lee's proposed method & 0.33 \\
Nik Defence & 0.43 - 0.45 \\
SDTLA & 0.47 \\
WVBM & Like Ideal scenario \\
\bottomrule
\end{tabular}
\end{table}

\subsubsection{The second experiment}
The objective of this experiment is to assess the effectiveness of the proposed models in preventing double-spending attacks. The section presents a graphical representation of the number of successful double-spending attacks based on the attacker's hash rate. The evaluation is conducted in a scenario where the network is equipped with the proposed methods, and the results are compared to other scenarios, such as a network that just uses a fixed $Z$ or a network equipped with a Nik-defense system that has been improved by using a fixed $Z$ to reduce double-spending attacks.

The table presented as \autoref{tbl1} displays the parameter values of the proposed systems used in this experiment. Notably, WVBM offers the key advantage of being able to determine the optimal value for $Z$. In the best-case scenario, $Z$ is equal to 2, while in the worst-case scenario, it is equal to 12. In comparison, SDTLA performs differently, with the $Z$ being 3 in the best cases and 24 in the worst cases.
\begin{table}[h]
\centering
\caption{Parameters used in the second experiment.}\label{tbl1}
\begin{tabular}{ccccc}

\toprule
Model & $Z$ interval & $K$ interval & $\tau$ & Time window \\
\midrule
SDTLA & 3--24 & 1--3 & 5 blocks & $12 \times \tau$ \\
WVBM & 2--12 & --- & 5 blocks & $12 \times \tau$ \\
\bottomrule

\end{tabular}
\end{table}

The following figure show the results of the second experiment for our proposed methods. In this figure, we can see the effects of our methods compared to previous works.
\begin{figure}
	\centering
		\includegraphics[scale=.60]{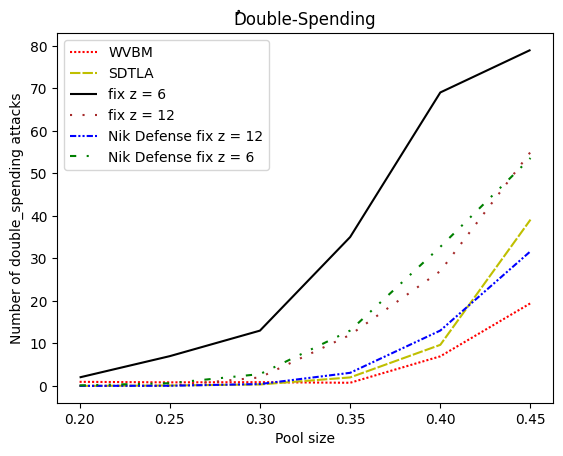}
	\caption{The second experiment, Compare our methods with previous works.}
	\label{FIG:3}
\end{figure}

As we can see in \autoref{FIG:3}, SDTLA can handle DS attacks better than improved Nik-Defense with a fixed $Z$. The $Z$ parameter in this method is often set between 6 and 12. So we can conclude that if we use this method, not only can we verify transactions faster than when we use Nik-Defense, but we can also reduce the risk of DS attacks better too.

Aside from the substantial reduction of double-spending attacks, it is crucial to consider the value of the parameter $Z$. Customers typically prefer prompt services, so increasing the value of $Z$ may not be favorable for them. In Bitcoin, a block is mined every ten minutes, implying that a rise in $Z$ from 6 to 12 would result in a 60-minute extension in service time.

According to the results, WVBM demonstrates superior performance in managing DS attacks compared to SDTLA and Nik-Defense. This technique sets the maximum allowable value for $Z$ at 12, but it frequently adjusts this parameter to 6, 4, or even 2. The effectiveness of the proposed methods can be evaluated by examining the average values of the $Z$ parameter and the time required for the service provider to dispatch goods, as shown in \autoref{tbl11}. Therefore, we can assert that WVBM is the best method proposed in this paper. SVBM uses a high $Z$ parameter to reduce the likelihood of double-spending attacks, but this approach may not be appropriate for services that demand quick delivery.
\begin{table}[h]
\centering
\caption{Average of Z parameter in proposed methods.}
\label{tbl11}
\begin{tabular}{ccc}
\toprule
Model & Average ($Z$) & Hours to Wait \\
\midrule
SDTLA & 16.54 & 2.756 \\
WVBM & 7.47 & 1.245 \\
\bottomrule
\end{tabular}
\end{table}

\subsubsection{The third experiment}
The objective of this experiment is to assess the impact of variations in tame intervals $\tau$ and Time Window on the performance of the two methods presented in this paper. As these parameters are directly associated with the learning automata employed in the proposed methods, the evaluation for SDTLA will encompass measuring its performance against double-spending and selfish mining attacks. On the other hand, for WVBM, the evaluation will focus solely on its performance against double-spending attacks.
\begin{itemize}
    \item \textbf{The impact of the $\tau$ time interval on proposed methods.}
    
Based on \autoref{FIG:4}, it is evident that the impact of this parameter on the SDTLA model's ability to address selfish mining is not particularly significant. It can be concluded that a value of $5$ for this parameter yields the most favorable outcomes. The rationale behind this is to enhance responsiveness in adapting to changes in the attacker's strength. However, this principle does not hold true for double-spending attacks. \autoref{FIG:5} and \autoref{FIG:6} clearly demonstrate that a higher value for this parameter yields superior results in both methods. Nonetheless, it is important to note that assigning a larger value to this parameter does not automatically imply its superiority.

In the conducted experiments, \autoref{tble3_1} and \autoref{tble3_2} present the values of the $Z$ parameter. Based on the information provided in this table, the optimal scenario occurs when $\tau$ is set to $5$. Considering the significance of transaction confirmation speed for both service providers and receivers, we designate $5$ as the appropriate value for this parameter. It is worth noting that the table's results indicate that when the model responds to the attacker's changes at a slower pace, larger values are assigned to $Z$. Although this reduces the risk of attacks, it also diminishes the transaction confirmation speed, which is undesirable for clients.

\begin{figure}
	\centering
		\includegraphics[scale=.60]{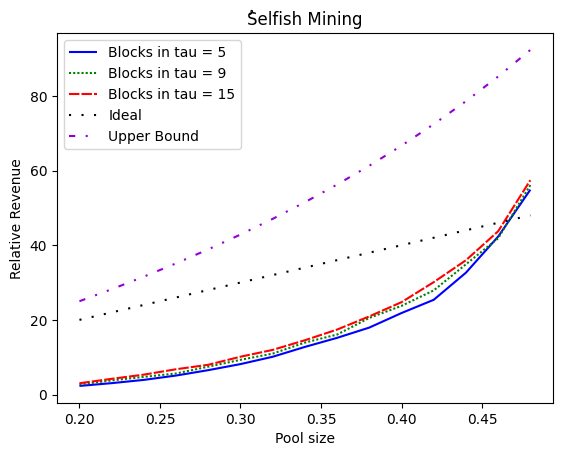}
	\caption{ The effect of the $\tau$ time interval on SDTLA in countering selfish mining.}
	\label{FIG:4}
\end{figure}

\begin{figure}
	\centering
		\includegraphics[scale=.60]{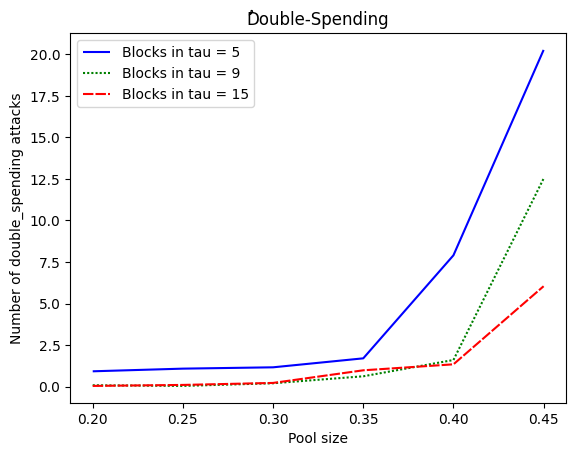}
	\caption{The effect of the $\tau$ time interval on WVBM in countering Double-spending attacks.}
	\label{FIG:5}
\end{figure}

\begin{figure}
	\centering
		\includegraphics[scale=.60]{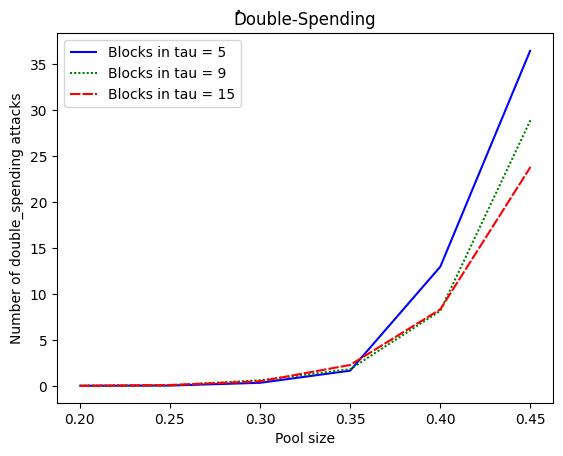}
	\caption{The effect of the $\tau$ time interval on SDTLA in countering Double-spending attacks.}
	\label{FIG:6}
\end{figure}

\begin{table}[h]
\centering
\caption{The effect of the $\tau$ time interval on SDTLA}\label{tble3_1}
\begin{tabular}{ccc}
\toprule
$\tau$ interval & Average ($Z$) & Hours to wait \\
\midrule
5 blocks & 16.54 & 2.756 \\
9 blocks & 19.16 & 3.193 \\
15 blocks & 19.38 & 3.23 \\
\bottomrule
\end{tabular}
\end{table}

\begin{table}[h]
\centering
\caption{The effect of the $\tau$ time interval on WVBM}\label{tble3_2}
\begin{tabular}{ccc}
\toprule
$\tau$ interval & Average ($Z$) & Hours to wait \\
\midrule
5 blocks    & 7.47  & 1.245 \\
9 blocks    & 9.56  & 1.593 \\
15 blocks   & 9.76  & 1.626 \\
\bottomrule
\end{tabular}
\end{table}

    \item \textbf{The impact of the $Time-Window$ time interval on proposed methods.}
    
According to the findings depicted in \autoref{FIG:7}, it becomes apparent that the influence of this parameter on the SDTLA model's capacity to combat selfish mining is relatively insignificant. Therefore, it can be inferred that employing a Time-Window size of $6 * \tau$ produces the most advantageous results. The underlying reasoning behind this choice is to enhance the system's ability to promptly adapt to variations in the attacker's capabilities. However, this principle does not hold true for double-spending attacks. 
\autoref{FIG:8} and \autoref{tble3_4} clearly demonstrate the superior performance of WVBM in mitigating double-spending attacks. Although utilizing a time window size of $6 * \tau$ yields exceptional results in terms of $Z$-mean, it also raises the possibility of facing weaker attackers compared to a window size of $12 * \tau$. Upon careful examination of \autoref{FIG:8}, \autoref{FIG:9}, \autoref{tble3_3}, and \autoref{tble3_4}, it becomes evident that opting for a Time-Window size of $12 * \tau$ would be the most prudent decision for both proposed methods. This choice effectively minimizes the risk of attacks to an appropriate level without excessively inflating the value of $Z$.

\begin{figure}
	\centering
		\includegraphics[scale=.60]{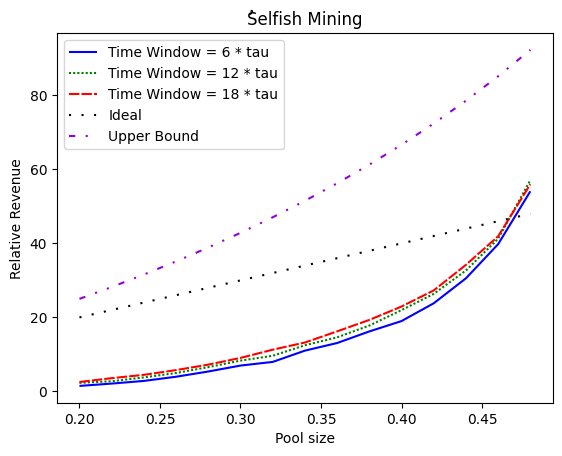}
	\caption{ The effect of the $Time-Window$ time interval on SDTLA in countering selfish mining.}
	\label{FIG:7}
\end{figure}

\begin{figure}
	\centering
		\includegraphics[scale=.60]{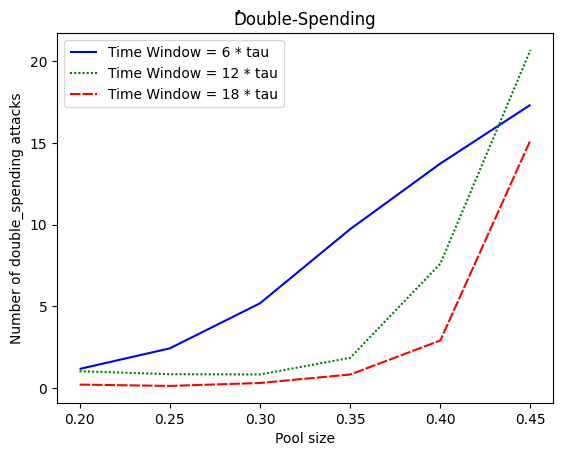}
	\caption{The effect of the $Time-Window$ time interval on WVBM in countering Double-spending attacks.}
	\label{FIG:8}
\end{figure}

\begin{figure}
	\centering
		\includegraphics[scale=.60]{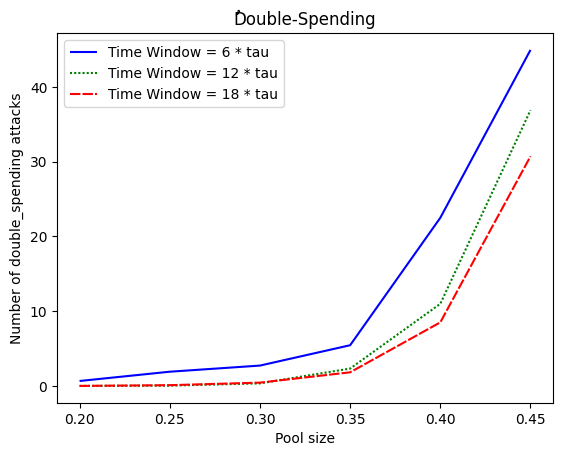}
	\caption{The effect of the $Time-Window$ time interval on SDTLA in countering Double-spending attacks.}
	\label{FIG:9}
\end{figure}

\begin{table}[h]
\centering
\caption{The effect of the Time-Window time interval on SDTLA}
\label{tble3_3}
\begin{tabular}{ccc}
\toprule
Size of Time-Window & Average ($Z$) & Hours to wait \\
\midrule
$6 \times \tau$ & 11.50 & 1.916 \\
$12 \times \tau$ & 16.59 & 2.765 \\
$18 \times \tau$ & 18.83 & 3.138 \\
\bottomrule
\end{tabular}
\end{table}

\begin{table}[h]
\centering
\caption{The effect of the \textit{Time-Window} time interval on WVBM}\label{tble3_4}
\begin{tabular}{ccc}
\toprule
Size of Time-Window & Average($Z$) & Hours to wait \\
\midrule
$6 \times \tau$    & 5.238  & 0.873 \\
$12 \times \tau$   & 7.51   & 1.251 \\
$18 \times \tau$   & 9.324  & 1.554 \\
\bottomrule
\end{tabular}
\end{table}
\end{itemize}
\section{Conclusion}
In this section, based on the findings from the experiments, we will present a comprehensive conclusion to assess the performance of the two proposed methods.
\subsection{SDTLA}
The experimental results indicate that SDTLA algorithm has a significant impact on reducing the risks of double-spending attacks and selfish mining. The algorithm outperforms Nik-defense system and is trained effectively due to the independence of the two learning automata. The results demonstrate the power of the learning automata in setting safe parameters and the intelligent performance of the algorithm in response to network changes. Compared to previous systems, SDTLA has a lower computational cost and higher ability to reduce the risk of attacks.

One important consideration in SDTLA is the possibility of the learning automata being stuck in a state where it repeatedly chooses actions with high probabilities. To address this issue, the learning automata are reset after a period of time and the parameters are returned to their initial values. While the main feature of VDHLA is the dynamic adjustment of the maximum depth for operation, an additional time interval is included to completely eliminate the risk of being stuck in a repeating state. The duration of this interval is chosen to be less than the time in which the attacker's processing power rate changes. It is essential to carefully set this parameter to avoid problems such as always choosing large values for Z. Another solution to this problem is to not provide rewards to the learning automata if they repeatedly choose the same actions. This approach ensures that the probability of choosing a particular action does not exceed a reasonable limit.

In summary, the advantages of SDTLA include its ability to effectively reduce double-spending attacks, its lower computational cost, and its intelligent response to network changes. The potential disadvantage of the algorithm is the need for careful parameter setting to avoid getting stuck in repeating states. However, solutions such as resetting the learning automata and not providing rewards for repeated actions effectively mitigate this issue.
Here are the advantages and disadvantages of this system.
\subsubsection{Advantages}
\begin{itemize}
    \item Preservation of the advantages of previous systems
    \item Simultaneous training of two learning automata with no adverse effects on each other
    \item Mitigation of the risk of selecting a winning chain in the presence of an eclipse attack
\end{itemize}
\subsubsection{Disadvantages}
\begin{itemize}
    \item Choosing the winning chain in a fork based on weight criteria when it's not necessary.
    \item Choosing a high value for the Z parameter, that result in slower transaction and trade processes.
    \item The potential for new attack methods to arise.

\end{itemize}
The meaning of the possibility of creating new attacks is that in this method, the choice of weight criteria plays a crucial role in most cases and this can be dangerous. For instance, if the attacker is two blocks behind and instead of relinquishing, they mine a block, an incorrect setting of the safe parameter associated with selfish mining and a poor selection of the winning chain based on the weight criterion can lead to the attacker winning. This situation can cause the emergence of new types of attacks, including selfish and double-spending attacks. To address this issue, the proposed solution is to design a system that can prevent these attacks. Therefore, we propose the WVBM system.
\subsection{WVBM}
Based on the experimental results, our findings indicate that the proposed solution is highly effective in mitigating selfish mining attacks. Despite a lower profitability threshold in some cases compared to SDTLA, the overall profitability for the attacker is significantly reduced, making it comparable to the ideal scenario which means no individual would be able to earn revenue greater than their proportionate processing power and as a result, there is no incentive for rational miners to act selfishly. While there is no defense that can achieve the conditions of the ideal scenario, WVBM can be considered the best outcome of this paper. It is worth noting that the success of the double-spending attack is highly dependent on the success of the selfish mining attack, and the use of WVBM has shown a considerable reduction in the possibility of such attacks, as evidenced by the experimental results. To provide a comprehensive evaluation of this system, we also present its advantages and disadvantages in the following sections.
\subsubsection{Advantages}
\begin{itemize}
    \item Preservation of the advantages of previous systems.
    \item Precise adjustment of the $Z$ parameter.
    \item Optimal defense performance against selfish mining.
    \item Reduced likelihood of selecting a winning chain during an eclipse attack.
    \item Elimination of potential vulnerabilities that could lead to new attacks.
    \item Generally, the length criterion is preferred, while the weight criterion is considered only in exceptional circumstances.
\end{itemize}
\subsubsection{Disadvantage}
\begin{itemize}
    \item The potential for new attack methods to arise.
\end{itemize}

\section{Summary and Future works}
Double-spending attacks are one of the most significant risks that digital currencies face. The risk of this type of attack increases when combined with selfish mining, so by preventing selfish mining, the risk of another attack can be significantly reduced. This paper introduced two intelligent, back-compatible decentralized defenses that use a new weight policy to select the winning chain. We propose two defensive systems to reduce the risks of selfish mining and double-spending attacks. We use learning automata as a light and fast learning tool to help our systems detect the possibility of these attacks. The results of experiments show that our models can reduce the risks of these kinds of attacks, and we can use them to improve the current blockchains. 

The SDTLA and WVBM methods have been proposed as effective defense mechanisms against selfish mining in blockchain networks. The SDTLA method has been shown to increase the profitability threshold of selfish mining up to 47$\%$, while the WVBM method performs even better and is very close to the ideal scenario where each miner's revenue is proportional to their shared hash processing power. Moreover, both methods can effectively mitigate the risks of double-spending through the tuning of the $Z$ Parameter. These findings highlight the potential of SDTLA and WVBM as promising solutions for enhancing the security and efficiency of blockchain networks.

Because of the weighting method used in defense systems presented in this paper, there will never be conditions under which two chains in a fork are completely equal in terms of weight. We know that normally when the lengths of two chains are equal, the winner will be the chain on which the next block is mined, which can lead to the attacker increasing his power by using the eclipse attack. Therefore, it can be said that the methods proposed in this paper have been able to eliminate the situation in which the attacker can use the eclipse attack to strengthen his combined attack.

Future research could explore the potential for new attacks arising from changes in system policies. One potential way for investigation is designing an attack where an automaton can switch between honest and selfish behavior at any moment. Such an attack could potentially disrupt the proper training of defense system automata and catch them off guard during the attack.

As blockchain technology continues to evolve, there is an increasing need for in-depth exploration and analysis of its potential applications. Notably, the incorporation of artificial intelligence and machine learning into blockchain technology has emerged as a fascinating area for further research. Therefore, the following recommendations are proposed for future studies in this domain:
\begin{enumerate}
    \item Investigating the efficacy of alternative reinforcement learning algorithms in mitigating blockchain attacks.
    \item Developing novel attack strategies utilizing learning automata to dynamically optimize the attacker's performance and response to defensive mechanisms.
    \item To enhance the security and robustness of distributed systems, future research could focus on developing advanced consensus algorithms based on machine learning techniques. These algorithms should be designed to make it significantly more challenging and costly for attackers to launch successful attacks against the system.
    \item Further work could be done to improve the efficiency and effectiveness of learning automata algorithms used in the proposed systems. This could involve developing more intelligent approaches for calculating the reinforcement signal that enables these algorithms to learn and adapt to changing environments. Such advancements could enhance the performance and reliability of these systems, leading to more widespread adoption and application in real-world scenarios.
    \item Develop a new algorithm for calculating the weight of chains to improve accuracy and efficiency.
    \item Investigate additional strategies to combat the combined attack of selfish mining, eclipse, and double-spending, which was proposed by Gervais et al.\cite{R21}, such as exploring alternative consensus mechanisms or implementing additional security measures.
    \item One potential area of future work could be the exploration of an intelligent approach for determining the optimal number of blocks used in the weighting of chains, based on network conditions. This approach would aim to improve the efficiency and effectiveness of the chain weighting process by leveraging intelligent algorithms to identify the optimal number of blocks for a given set of network conditions. Such an approach could potentially enhance the scalability and performance of blockchain networks, and warrants further investigation in future research.
    \item An area for future exploration involves leveraging reinforcement learning algorithms, such as learning automata, to intelligently calculate the chain length coefficient for obtaining a weighted threshold in WVBM. This approach has the potential to improve the accuracy and efficiency of calculations
\end{enumerate}

\bibliographystyle{unsrtnat}

\bibliography{cas-refs}

\begin{IEEEbiography}[{\includegraphics[width=1in,height=1.25in,clip,keepaspectratio]{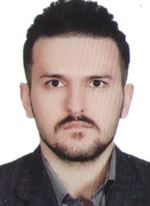}}]{Seyed Ardalan Ghoreishi}
He earned his Bachelor of Science in Electrical Engineering from Sadjad University of Technology in 2014. In 2022, he earned his Master of Science in Computer Engineering from Amirkabir University of Technology in Tehran, Iran. His areas of expertise include Machine Learning, Deep Learning, and Blockchain technology.
\end{IEEEbiography}

\begin{IEEEbiography}[{\includegraphics[width=1in,height=1.25in,clip,keepaspectratio]{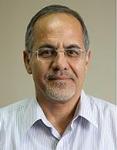}}]{Mohammad Reza Meybodi}
He earned his Bachelor of Science and Master of Science in Economics from Shahid Beheshti University in Tehran, Iran in 1973 and 1977 respectively. He later received his Master of Science and Doctorate in Computer Science from Oklahoma University in Norman, OK, USA in 1980 and 1983 respectively. Afterward, he served as an Assistant Professor at Western Michigan University in Kalamazoo, MI from 1983 to 1985, and then as an Associate Professor at Ohio University in Athens, OH from 1985 to 1991. Presently, he holds the position of a Full Professor at the Computer Engineering Department of Amirkabir University of Technology in Tehran. His areas of expertise encompass Wireless Networks, Fault-Tolerant Systems, Learning Systems, Parallel Algorithms, Soft Computing, and Software Development.
\end{IEEEbiography}

\end{document}